\documentclass[fleqn,twoside]{article}
\usepackage{espcrc2}

\usepackage{graphicx}
\usepackage[figuresright]{rotating}

\newcommand{\AmS}{{\protect\the\textfont2
  A\kern-.1667em\lower.5ex\hbox{M}\kern-.125emS}}

\title{Effective Potential for Polyakov Loops in Lattice QCD}

\author{Y. Nemoto\address[RBC]{RIKEN BNL Research Center,
        Brookhaven National Laboratory, NY 11973, USA}
	[RBC collaboration]
	\thanks{We thank RIKEN, Brookhaven National Laboratory and the U.S.\
	Department of Energy for providing the facilities essential for the
	completion of this work.}
	}
       
\begin{document}

\begin{abstract}
Toward the derivation of an effective theory for Polyakov loops
in lattice QCD, we examine Polyakov loop correlation functions using 
the multi-level algorithm which was recently developed by Luscher 
and Weisz.
\vspace{1pc}
\end{abstract}

\maketitle

\section{Introduction}

Precise measurement of Polyakov loop correlation is important for 
finite temperature physics.
First of all heavy quark potentials at finite temperature are
obtained by measuring two-Polyakov-loop correlation functions.
It is already well-known that we can obtain the $Q\bar{Q}$ 
potential below and above the critical temperature ($T_C$)
from this measurement.
Due to the nonperturbative effects, however, there
are still less-known quantities at intermediate temperature.
For example the $Q\bar{Q}$ potential at finite temperature
is generally written as $V/T = -e/(RT)^d \exp(-\mu R)$,
where $T$ is temperature, $R$ the distance between quarks, and
$e, d, \mu$ are parameters.
At very high temperature, perturbation theory predicts $d=2$,
while just above $T_C$, lattice simulation gives the deviation from
$d=2$\cite{Kac00}.
In relation to such a nonperturbative behavior at finite temperature,
our main interest here is to derive an effective
theory for Polyakov loops from lattice QCD.
Recently new Polyakov loop model is proposed\cite{Pis02}
regarding the deconfining phase transition in $SU(N)$ gauge theory.
It conjectures a relationship between the Polyakov loops and
the pressure.
The usual Polyakov loop $\ell_1$ transforms under a global 
$Z(N)$ transformation as $\ell_1 \to e^{i\phi}\ell_1$, with
$\phi=2\pi j/N (j=1...N-1)$.
This means that $\ell_1$ transforms as a field with charge one.
On the other hand one can consider Polyakov loops with
higher charge under a global $Z(N)$ transformation.
For example the charge two Polyakov loop is defined as
\begin{equation}
  \ell_2 = \frac{1}{N}{\rm tr} L^2 
           - \frac{1}{N^2}({\rm tr} L)^2,
\end{equation}
where $L$ is an ordered link product in the temporal direction,
$L(x)={\rm P}\exp(ig\int_0^\beta A_0(x,\tau)d\tau)$.
The Polyakov loops with higher charge might affect non-universal
behavior in the model.
Therefore it is interesting to measure relevant kind of new
Polyakov loop correlation on the lattice without any
phenomenological assumptions.
In this case, however, we need the Polyakov loops with
relatively large length in the temporal direction,
which might be inaccessible in the ordinary computation 
of the Polyakov loops.
Thus we are led to try the new algorithm recently 
developed by Luscher and Weisz called the multi-level method\cite{LW01}
to measure various Polyakov loop correlation functions.
This method can measure quite precisely for the 
Polyakov loop correlation with the large temporal length
and/or large distance between the Polyakov loops
as shown below.

\section{Precise measurement of Polyakov loop correlation}

One way to get the Polyakov loop correlation functions
as precisely as possible
is the method of improved estimator.
When we measure Polyakov loops, instead of a time-like link 
variable itself, by using some average of the time-like links,
we can reduce the error of the Polyakov loop without affecting
the average.
The multi-level algorithm, which is used here, is more
efficient in measuring the Polyakov loop correlation.
We explain this algorithm briefly in the following.
Let us consider the two-Polyakov-loop correlation,
$\langle P(n) P^\dagger(n+r) \rangle$,
where $n$ is a spatial coordinate and $r$ is the distance
between two quarks. 
First we define a two-time-like link operator, 
\begin{equation}
  T(t_0,r)_{\alpha \beta \gamma \delta} = U_4(t_0,n)_{\alpha \beta}
  U_4^\dagger(t_0,n+r)_{\gamma \delta}.
\end{equation}
The essence of the multi-level algorithm is similar to the multi-hit
one, i.e., is to take an average of $T$ instead of $T$ itself
in the sublattice of the time slice,
\begin{equation}
  [ T(t_0,r)_{\alpha \beta \gamma \delta} ] = \frac{1}{Z_s}
  \int D[U]_s T(t_0,r) e^{-S[U]_s}.
\end{equation}
This is schematically shown in Fig.1.
Here $s$ denotes the sublattice which includes
the operator $T$ and is shown as the shaded area 
in Fig.1.
\begin{figure}[t]
\vspace{9pt}
\begin{center}
\includegraphics[width=2.2cm]{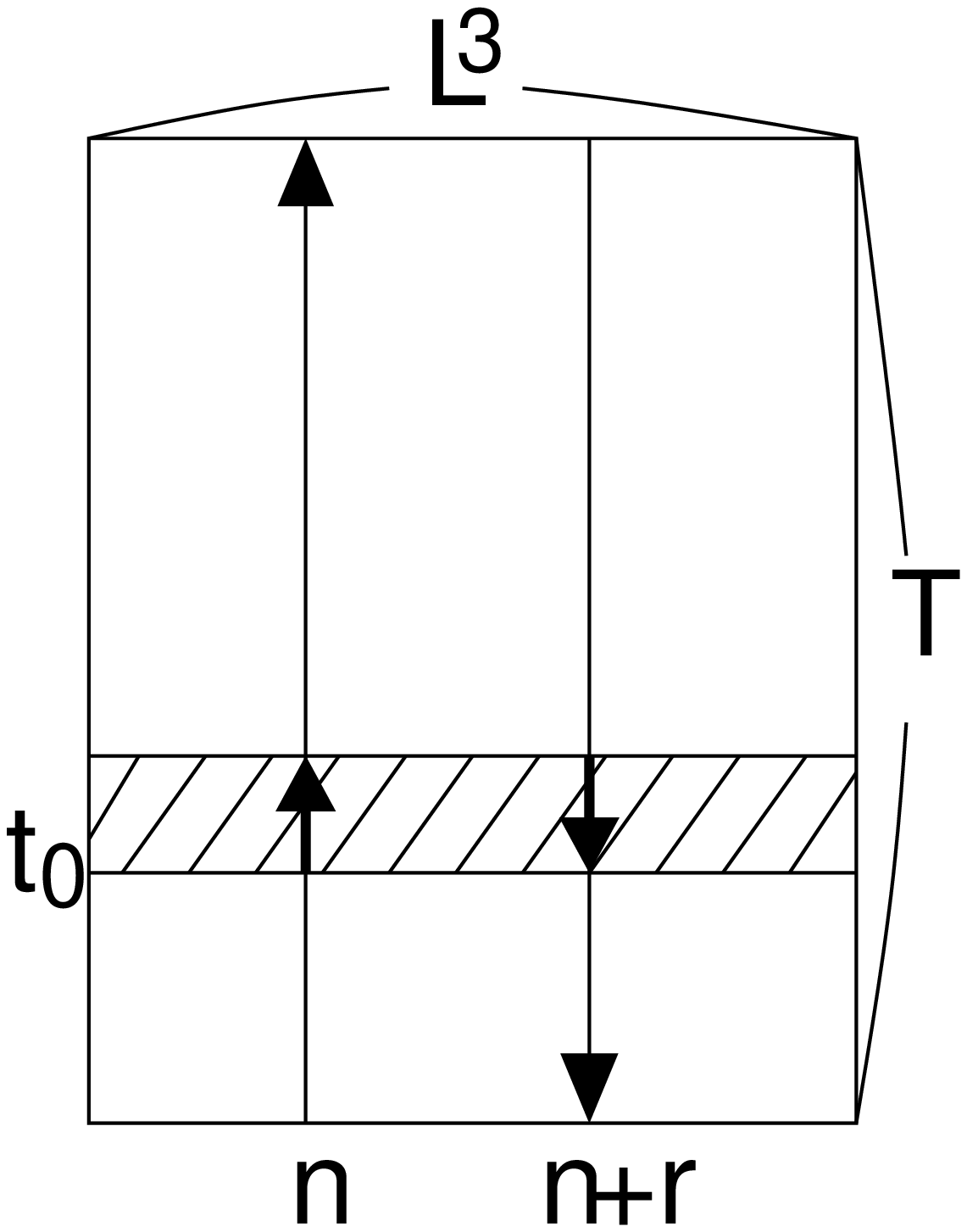}
\hspace{1cm}
\includegraphics[width=3cm]{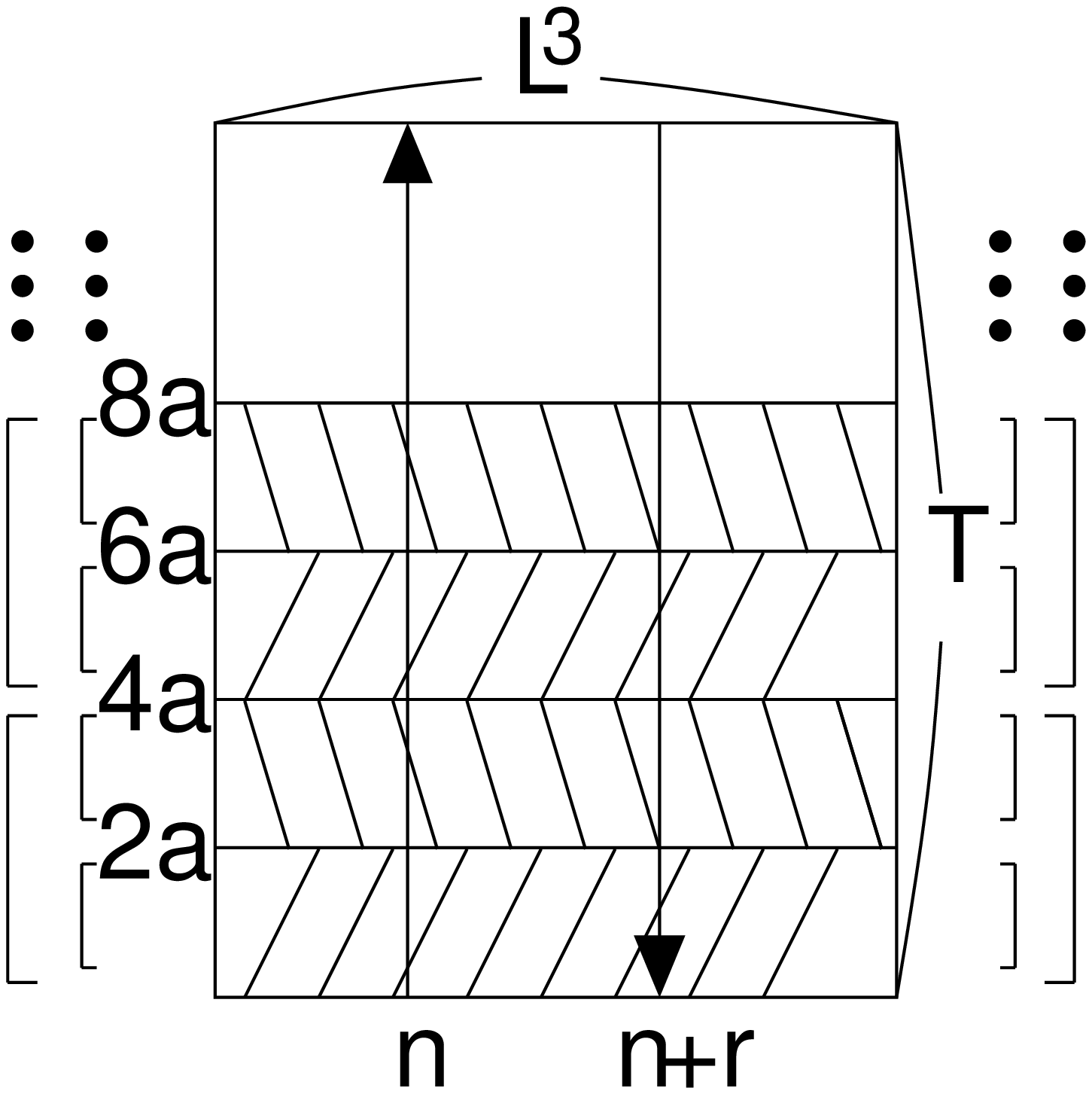}
\caption{A two-time-like link operator (denoted by thick lines, left
figure) and two-Polyakov-loop correlation with the multi-level algorithm,
eqs.(\ref{eq:onel}) or (\ref{eq:twol}) (right).
}
\end{center}
\label{fig:twotime}
\end{figure}
The Polyakov loop correlation function is then expressed as
\begin{eqnarray}
  \lefteqn{\langle P(n)P^\dagger(n+r)\rangle} \nonumber \\
  &=& \langle [T(0,r)T(a,r)][T(2a,r)T(3a,r)] \nonumber \\
  & & \cdots [T(T-2a,r)T(T-a,r)] \rangle
  \label{eq:onel}
\end{eqnarray}
for single layer or
\begin{eqnarray}
  \lefteqn{\langle P(n)P^\dagger(n+r)\rangle} \nonumber \\
  &=& \langle \left[ [T(0,r)T(a,r)][T(2a,r)T(3a,r)] \right] \nonumber \\
  & & \qquad \cdots \left[ \cdots[T(T-2a,r)T(T-a,r)]\right] \rangle 
  \label{eq:twol}
\end{eqnarray}
for two layers.
These equations are also schematically shown in Fig.1.

In this method there are essentially two new parameters, which we call
$n_{\rm ms}$ and $n_{\rm up}$.
$n_{\rm ms}$ is the number of measurement to be made for the time slice
average
\begin{eqnarray}
  \lefteqn{[T(t_0,r)T(t_0+a,r)]}\nonumber \\
  &=& \frac{1}{n_{\rm ms}}\sum_i^{n_{\rm ms}} T_i(t_0,r) T_i(t_0+a,r),
\end{eqnarray}
and $n_{\rm up}$ is the number of time slice updates between measurement.
We determine $n_{\rm ms}$ so that the s/n ratio is equal to unity.
Then the statistical errors are reduced exponentially as the distance
between the Polyakov loops increases\cite{LW01}.

We show some test results of the algorithm in the following.
All the simulations are done with the lattice size $12^4$ at
$\beta=5.7$.
We use the standard Wilson action and no gauge fixing.
Fig.2 shows the $n_{\rm ms}$ dependence of a 
two-Polyakov-loop correlation function at a certain gauge configuration.
In the figure,
the spatial distance between the two Polyakov loops is set to $4$
in the lattice unit and $n_{\rm up}=1$ is fixed.
In the single layer case, the relative computation time in the
horizontal line is proportional to $n_{ms}$, i.e.,
the leftmost point corresponds to $n_{\rm ms}=10$ and the rightmost 
 $n_{\rm ms}=1000$.
In the two layer case, we have chosen the parameters
$n_{\rm ms}=10$ and $n_{\rm up}=16$ for the outer layer and
$n_{\rm ms}=5\sim 150$ and $n_{\rm up}=1$ for the inner layer.
Usual measurement of the Polyakov loop correlation without
any improvement corresponds to $n_{\rm ms}=1$.
The graph shows that the values approach the precise one as
$n_{\rm ms}$ increases.
We have confirmed that our calculation agrees with that of
\cite{LW01} within the range of the statistical error.

We have also calculated three-Polyakov-loop correlation shown in
Fig.3.
As an example of the result we show the $n_{\rm ms}$ dependence of
the quantity $\langle |P(n+4x)P(n+4y)P(n+4z)|\rangle$
in Fig.4. 
We see the validity of the algorithm even in this case by increasing
the value of $n_{\rm ms}$.
 
\begin{figure}[t]
\vspace{9pt}
\begin{center}
\includegraphics[width=7cm]{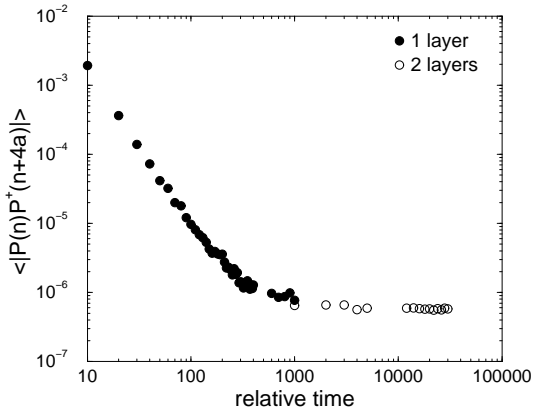}
\caption{Two-Polyakov-loop correlation with the distance $r=4a$.}
\end{center}
\label{fig:2hi}
\end{figure}
\begin{figure}[ht]
\vspace{9pt}
\begin{center}
\includegraphics[width=3cm]{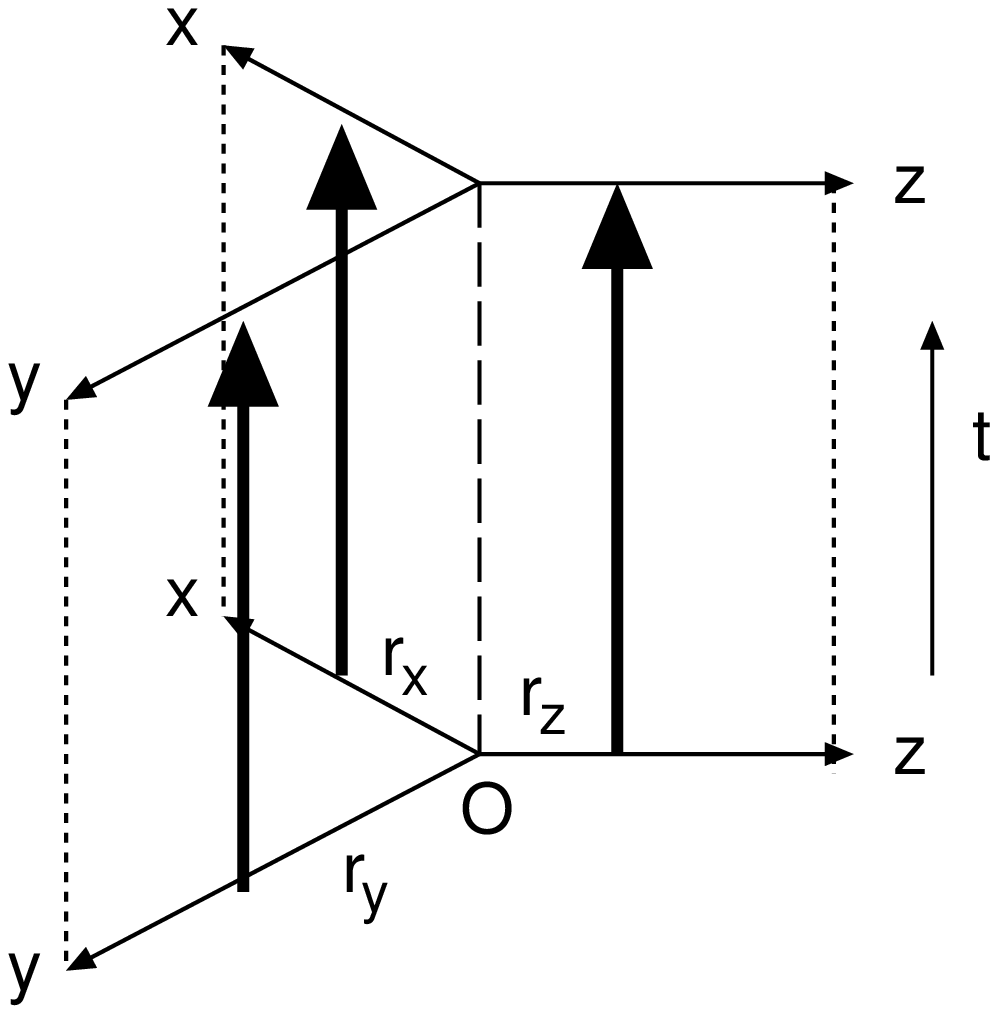}\\
\caption{Schematic figure of the three-Polyakov-loop correlation.}
\vspace{4mm}
\includegraphics[width=7cm]{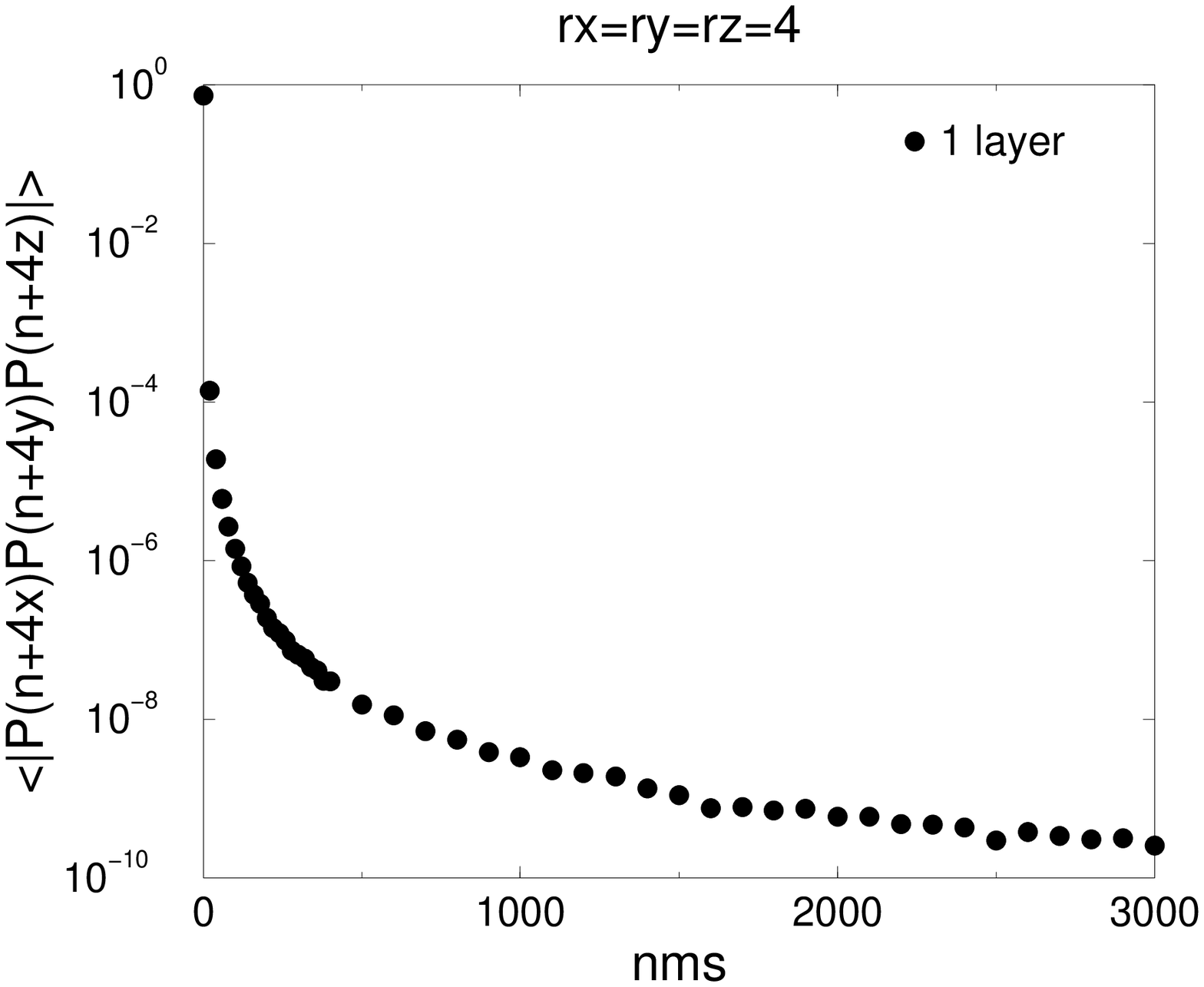}
\caption{$n_{\rm ms}$ dependence of the three-Polyakov-loop 
correlation function.}
\end{center}
\label{fig:3p1h}
\end{figure}

Finally we comment on our trial computation of a variant
of the multi-level method.
Measurement of many-point Polyakov loop correlation in the above method 
spends much memory.
One possible solution is to apply a method which is used in the
calculation of quark propagators of disconnected diagrams.
Let us consider for simplicity the two-Polyakov-loop correlation.
The modified method is to take summation over space of the two-time-link 
operators before taking the tensor product of them,
\begin{eqnarray}
  \lefteqn{\langle P(n)P^\dagger (n+r)\rangle
  = \frac{1}{N_{\rm conf}} 
  \{ [[K(0)][K(2a)]] } \nonumber \\
  & & \cdots[[K(T-4a)][K(T-2a)]]   \},\\
  &&K(t_0) = \sum_{\rm space} T(t_0,r)T(t_0+a,r).
\end{eqnarray}
If the number of configuration and $n_{\rm ms}$ are 
sufficiently large,
this method gives the same results as the conventional 
multi-level method,
because the gauge non-invariant terms cancel out.
Our trial simulation is, however, much noisier than the 
conventional one due to the finite
number of configuration and $n_{\rm ms}$.
So unfortunately our present results imply that this method is 
less effective.

In summary,
toward the construction of the effective theory for
Polyakov loops, we have tried a new algorithm called the
multi-level method\cite{LW01}.
We have reproduced the two-Polyakov-loop correlation by
\cite{LW01} and confirmed the validity for the computation
of the three-Polyakov-loop correlation.


\begin{thebibliography}{9}
\bibitem{Kac00} O. Kaczmarek, F. Karsch, E. Laermann and
     	M. L\"utgemeier, Phys. Rev. D {\bf 62} (2000) 034021.
\bibitem{Pis02} R.D. Pisarski, Nucl. Phy. A {\bf 702} (2002) 151;
                hep-ph/0203271; A. Dumituru and R.D. Pisarski,
                hep-ph/0204233.
\bibitem{LW01} M. Luscher and P. Weisz, JHEP {\bf 109} (2001) 010.
\end{thebibliography}
\end{document}